\documentclass[preprintnumbers,twocolumn,unsortedaddress,amsmath,prb,nobibnotes,altaffilletter]{revtex4}

\usepackage{graphics,graphicx}
\preprint{Physics of Fluids}

\begin{document}

\title{On Landau's prediction for large-scale fluctuation of turbulence energy dissipation}

\author{Hideaki Mouri}
\email{hmouri@mri-jma.go.jp}
\affiliation{Meteorological Research Institute, Nagamine, Tsukuba 305-0052, Japan}

\author{Masanori Takaoka}
\email{mtakaoka@mail.doshisha.ac.jp}
\affiliation{Department of Mechanical Engineering, Doshisha University, Kyotanabe, Kyoto 610-0321, Japan}

\author{Akihiro Hori}
\altaffiliation[Also at ]{Meteorological and Environmental Service, Inc.}

\author{Yoshihide Kawashima}
\altaffiliation[Also at ]{Meteorological and Environmental Service, Inc.}
\affiliation{Meteorological Research Institute, Nagamine, Tsukuba 305-0052, Japan}

\date{December 3, 2005}

\begin{abstract}
Kolmogorov's theory for turbulence in 1941 is based on a hypothesis that small-scale statistics are uniquely determined by the kinematic viscosity and the mean rate of energy dissipation. Landau remarked that the local rate of energy dissipation should fluctuate in space over large scales and hence should affect small-scale statistics. Experimentally, we confirm the significance of this large-scale fluctuation, which is comparable to the mean rate of energy dissipation at the typical scale for energy-containing eddies. The significance is independent of the Reynolds number and the configuration for turbulence production. With an increase of scale $r$ above the scale of largest energy-containing eddies, the fluctuation becomes to have the scaling $r^{-1/2}$ and becomes close to Gaussian. We also confirm that the large-scale fluctuation affects small-scale statistics.
\end{abstract}


\maketitle

\section{INTRODUCTION}
\label{s1}

For locally isotropic turbulence, Kolmogorov \cite{k41a} based his theory in 1941 on a hypothesis that small-scale statistics are uniquely determined by the kinematic viscosity $\nu$ and the mean rate of energy dissipation $\langle \varepsilon \rangle$. Landau \cite{ll59} remarked as follows. The local rate of energy dissipation $\varepsilon$ should fluctuate in space over scales of large eddies. This large-scale fluctuation should not be universal but should be different for different flows. Since the large-scale fluctuation should affect small-scale statistics, there should be no universal law that describes the small-scale statistics.

Landau's remark has been wrongly regarded as a prediction for small-scale intermittency, i.e., enhancement of a small-scale fluctuation in a fraction of the volume. This is because, when Kolmogorov \cite{k62} revised his theory in 1962 to incorporate small-scale intermittency, he gave too much credit to Landau.\cite{f91,f95} For small scales, Landau remarked not about intermittency but about universality.

The small-scale statistics in Landau's original remark are the second-order moments of velocity increments alone. Here we extend Landau's remark to consider the universality of other small-scale statistics.

The importance of Landau's remark is not only to small-scale statistics. Any potentially significant phenomenon at large scales is in itself important because the large scales dominate the turbulence energy.

Despite numerous studies of small-scale intermittency in the energy dissipation, \cite{k62,f91,f95,ms91,ss95,sa97,po97,kg02,c03} its large-scale fluctuation has not attracted much interest. Experimentally, the fluctuation has been known to exist on some level, but it has not been discussed with any highlighting. The existing works are all theoretical. Oboukhov \cite{o62} considered that the fluctuation is significant. Kraichnan \cite{k74} considered that the fluctuation is smoothed out by spatial mixing, which could be caused by the action of pressure fluctuation. \cite{ms91} Frisch \cite{f91,f95} considered that the fluctuation is significant in some specific configurations for turbulence production. Thus, even for the significance of the fluctuation, no consensus has been reached on.

We experimentally assess Landau's remark, i.e., whether the large-scale fluctuation of energy dissipation is significant, whether the large-scale fluctuation is universal, and whether the large-scale fluctuation affects small-scale statistics. The assessment requires more than one type of turbulence. We study grid and boundary-layer turbulence at several Reynolds numbers. With a hot-wire anemometer and Taylor's frozen-eddy hypothesis, we obtain a one-dimensional cut of the velocity field. The local rate of energy dissipation per unit mass is obtained along the streamwise position $x$ as
\begin{equation}
\varepsilon _u (x) = 15 \nu \left[ \frac{du(x)}{dx} \right] ^2
\
\mbox{or}
\quad
\varepsilon _v (x) = \frac{15 \nu}{2} \left[ \frac{dv(x)}{dx} \right] ^2 .
\end{equation}
Here $u$ and $v$ are the streamwise and transverse velocities. These rates are surrogates of the true rate $\nu  \sum _{i,j} ( \partial _i u_j + \partial _j u_i)^2 /2$, where $i$ and $j$ denote coordinate axes. Their averages are nevertheless exact if turbulence is locally isotropic. The energy dissipation at the scale $r$ is obtained by coarse-graining the local rate: \cite{o62}
\begin{equation}
\label{eq2}
\varepsilon (r,x) = \frac{1}{r} \int ^{+r/2}_{-r/2} \varepsilon (x+x') dx'
.
\end{equation}
For this and other equations, the subscript $u$ or $v$ is omitted if it is not necessary. The fluctuation of the energy dissipation $\varepsilon (r,x)$ around its average $\langle \varepsilon (r,x) \rangle = \langle \varepsilon (x) \rangle$ is studied as a function of scale $r$.

We study the features expected for largest scales in Sec. \ref{s2}. Grid turbulence is studied in Sec. \ref{s3}. Boundary-layer turbulence is studied in Sec. \ref{s4}. The dependence on the Reynolds number is studied in Sec. \ref{s5}. We summarize our conclusions and discuss the effect on small-scale statistics in Sec. \ref{s6}. Taylor's frozen-eddy hypothesis for large scales is studied in Appendix.

\begingroup
\squeezetable
\begin{table*}[!]
\caption{\label{t1} Summary of flow parameters.}
\begin{ruledtabular}
\begin{tabular}{lllccc}
\multicolumn{2}{l}{Quantity}                                                                                                                & Units            & Grid       &\multicolumn{2}{c}{Boundary layer}\\ 
                                                 &                                                                                          &                  &            & $z=0.25$\,m & $z=0.70$\,m \\
\cline{5-6}
\hline
 Mean streamwise velocity                        & $U$                                                                                      & m\,s$^{-1}$      & 10.57      & 7.05        & 10.22       \\
 Streamwise velocity fluctuation                 & $\langle u^2 \rangle ^{1/2}$                                                             & m\,s$^{-1}$      & 0.525      & 1.48        & 0.642       \\
 Spanwise velocity fluctuation                   & $\langle v^2 \rangle ^{1/2}$                                                             & m\,s$^{-1}$      & 0.515      & 1.22        & 0.544       \\
 Streamwise flatness factor                      & $\langle u^4 \rangle / \langle u^2 \rangle ^2$                                           &                  & 3.02       & 2.71        & 7.83        \\
 Spanwise flatness factor                        & $\langle v^4 \rangle / \langle v^2 \rangle ^2$                                           &                  & 2.99       & 3.02        & 7.36        \\ 
 Air temperature                                 &                                                                                          & $^{\circ}$C      & 11.6--12.0 & 14.9--15.9  & 14.4--14.9  \\
 Kinematic viscosity                             & $\nu$                                                                                    & cm$^2$\,s$^{-1}$ & 0.142      & 0.145       & 0.145       \\
 Mean energy dissipation rate ($\varepsilon _u$) & $\langle \varepsilon _u \rangle = 15 \nu \langle (\partial _x u )^2 \rangle$             & m$^2$\,s$^{-3}$  & 1.22       & 5.45        & 0.575       \\
 Mean energy dissipation rate ($\varepsilon _v$) & $\langle \varepsilon _v \rangle = 15 \nu \langle (\partial _x v )^2 \rangle /2$          & m$^2$\,s$^{-3}$  & 1.20       & 4.05        & 0.493       \\
 Correlation length ($u$)              & $L_u = \int ^{\infty}_0 \phi _u (r) dr / \phi _u (0)  $                                            & cm               & 17.4       & 42.3        & 35.9        \\
 Correlation length ($v$)              & $L_v = \int ^{\infty}_0 \phi _v (r) dr / \phi _v (0)  $                                            & cm               & 4.43       & 5.78        & 7.60        \\
 Correlation length ($\varepsilon _u$) & $L_{\varepsilon _u} = \int ^{\infty}_0 \phi _{\varepsilon _u} (r) dr / \phi _{\varepsilon _u} (0)$ & cm               & 0.545      & 0.970       & 2.69        \\
 Correlation length ($\varepsilon _v$) & $L_{\varepsilon _v} = \int ^{\infty}_0 \phi _{\varepsilon _v} (r) dr / \phi _{\varepsilon _v} (0)$ & cm               & 0.414      & 0.753       & 2.03        \\
 Taylor microscale                               & $\lambda = [ 2 \langle v^2 \rangle / \langle (\partial _x v )^2 \rangle ]^{1/2}$         & cm               & 0.688      & 0.896       & 1.14        \\
 Kolmogorov length                               & $\eta = (\nu ^3 / \langle \varepsilon _v \rangle )^{1/4}$                                & cm               & 0.0221     & 0.0166      & 0.0280      \\
 Microscale Reynolds number                      & Re$_{\lambda} = \langle v^2 \rangle ^{1/2} \lambda / \nu$                                &                  & 249        & 756         & 428         \\
\end{tabular}
\end{ruledtabular}
\end{table*}
\endgroup

\section{LARGEST-SCALE FEATURES}
\label{s2} 

Suppose that the local rate of energy dissipation $\varepsilon (x)$ is obtained on a one-dimensional cut of a turbulence flow. If the turbulence is homogeneous along the one-dimensional cut, the correlation function $\phi _{\varepsilon} (r)$ at the scale $r$ is 
\begin{equation}
\phi _{\varepsilon} (r) = \langle [ \varepsilon (x+r) - \langle \varepsilon (x) \rangle ]
                                   [ \varepsilon (x)   - \langle \varepsilon (x) \rangle ] 
                          \rangle 
.
\end{equation}
Here $\langle \cdot \rangle$ denotes a spatial average over the position $x$. The correlation length $L_{\varepsilon}$ is
\begin{equation}
L_{\varepsilon} = \frac{ \int ^{\infty}_0 \phi _{\varepsilon} (r) dr}
                       {                  \phi _{\varepsilon} (0)   }
.
\end{equation}
We assume that the correlation is insignificant or absent above a certain scale $r_{\ast}$, i.e., $\phi _{\varepsilon} (r) / \phi _{\varepsilon} (0) \simeq 0$ at $r > r_{\ast} > L_{\varepsilon}$. This assumption is natural for the correlation of any quantity in real turbulence that only has a finite extent.

The correlation function $\phi _{\varepsilon} (r)$ allows us to obtain the mean-square fluctuation of the energy dissipation $\varepsilon (r,x)$ around its average $\langle \varepsilon (r,x) \rangle$:
\begin{equation}
\sigma _{\varepsilon} (r)^2
=
\left\langle  [ \varepsilon (r,x) - \langle \varepsilon (r,x) \rangle ]^2 \right\rangle
=
\frac{2}{r^2} \int ^r_0 (r-r') \phi _{\varepsilon} (r') dr' 
.
\end{equation}
If the scale $r$ is much greater than the scale $r_{\ast}$, we have
\begin{equation}
\sigma _{\varepsilon} (r)^2
=
\frac{2L_{\varepsilon} \phi _{\varepsilon} (0)}{r}
.
\end{equation}
Thus the root-mean-square fluctuation scales as $\sigma _{\varepsilon} (r) \propto r^{-1/2}$.

We explain the scaling $\sigma _{\varepsilon} (r) \propto r^{-1/2}$. If the scale $r$ is much greater than the scale $r_{\ast}$, the energy dissipation $\varepsilon (r,x)$ is written with the sum over subregions that have a width $r_{\ast}$:
\begin{equation}
\label{eq7}
\varepsilon (r,x) = \frac{r_{\ast}}{r} 
                    \sum _{n=1}^{r/r_{\ast}} \varepsilon (r_{\ast},x_n),
\end{equation}
where $x_n = x -(r+r_{\ast})/2 + nr_{\ast}$. Since the energy dissipations $\varepsilon (r_{\ast},x_n)$ are mutually independent, the mean-square fluctuation for $\sum _n \varepsilon (r_{\ast},x_n)$ is $\sum _n \sigma _{\varepsilon} (r_{\ast})^2$. Then we have
\begin{equation}
\sigma _{\varepsilon} (r)^2 = \frac{r_{\ast} \sigma _{\varepsilon} (r_{\ast})^2}{r},
\end{equation}
which again leads to the scaling $\sigma _{\varepsilon} (r) \propto r^{-1/2}$. Thus this scaling is not dynamical but statistical.

There is another largest-scale statistical feature. The total number of subregions $r/r_{\ast}$ in Eq. (\ref{eq7}) is large. Then the central limit theorem ensures that the energy dissipation $\varepsilon (r,x)$ obeys a Gaussian distribution at least in the vicinity of the average $\langle \varepsilon (r,x) \rangle$, \cite{kg02} provided that the fluctuation of energy dissipation $\sigma _{\varepsilon} (r)$ is sufficiently smaller than the average $\langle \varepsilon (r,x) \rangle$.

The scaling $\sigma _{\varepsilon} (r) \propto r^{-1/2}$ and the tendency for Gaussianity imply that $r \varepsilon (r,x)$ is additive at largest scales. It is not additive at the smaller scales. The value of an additive quantity for a region is the sum of its values for the subregions that are mutually independent [Eq. (\ref{eq7})]. Many examples exist in the statistical mechanics and thermodynamics. \cite{ll79} There the scales of interest are much greater than the scales where the correlation is significant.

\section{GRID TURBULENCE}
\label{s3}

\subsection{Experiment}
\label{s3a}

The experiment was done in a wind tunnel of the Meteorological Research Institute. We use the coordinates $x$, $y$, $z$ in the streamwise, spanwise, and floor-normal directions. The corresponding turbulence velocities are $u$, $v$, and $w$. The origin $x = y = z = 0$\,m is taken on the tunnel floor at the entrance to the test section. Its size was $\delta x = 18$\,m, $\delta y = 3$\,m, and $\delta z = 2$\,m.

We placed a grid across the entrance to the test section at $x = 0$\,m. The grid consisted of two layers of uniformly spaced rods, the axes of which were perpendicular to each other. The separation of the axes of adjacent rods was 0.20\,m. The cross section of the rods was $0.04 \times 0.04$\,m. We set the mean wind to be $U \simeq 10$\,m\,s$^{-1}$.

The streamwise ($u$) and spanwise ($v$) velocities were simultaneously measured using a hot-wire anemometer. The anemometer was composed of a constant temperature system and a crossed-wire probe. The wires were made of platinum-plated tungsten, 5\,$\mu$m in diameter, 1.25\,mm in sensitive length, 1\,mm in separation, and 280\,$^{\circ}$C in temperature.

The measurement was done on the tunnel centerline at $x = 4$\,m, where the flatness factors $\langle u(x)^4 \rangle / \langle u(x)^2 \rangle ^2$ and $\langle v(x)^4 \rangle / \langle v(x)^2 \rangle ^2$ were close to the Gaussian value of 3. Thus turbulence was well developed, so eddies with various sizes and strengths filled the space randomly and independently. \cite{m02,m03a} If the measurement position had been too close to or far from the grid, turbulence should have been still developing or already decaying, and its flatness factors should have been different from the Gaussian value. The ratio $\langle u^2 \rangle ^{1/2} / \langle v^2 \rangle ^{1/2}$ was close to unity.

The signal was linearized, low-pass filtered at 15 kHz with 24 dB per octave, and then sampled digitally at 30 kHz with 16-bit resolution. The total length of the signal was $3 \times 10^8$ points.

In addition, to study the average and fluctuation of energy injection at $x = 4$\,m, the streamwise velocity was simultaneously measured at $x = 3.75$ and 4.25\,m. We used two single-wire probes. The total length of the signal was $10^8$ points. The other conditions were the same as in the above measurement.

The flow parameters at $x = 4$\,m are summarized in Table \ref{t1}. We used Taylor's frozen-eddy hypothesis to convert temporal variations into spatial variations in the streamwise direction (see Appendix). The velocity derivatives were obtained from finite differences, e.g., 
\begin{equation}
\frac{d u(x)}{d x} = \frac{u(x+\delta x)-u(x-\delta x)}{2\delta x}.
\end{equation}
Here $\delta x$ is the sampling interval. Since the sampling interval was small enough, the higher-order accuracy for the velocity derivatives is not necessary. \cite{ms91}

\begin{figure}[!]
\resizebox{8cm}{!}{\includegraphics*[4.5cm,7.5cm][19.5cm,26cm]{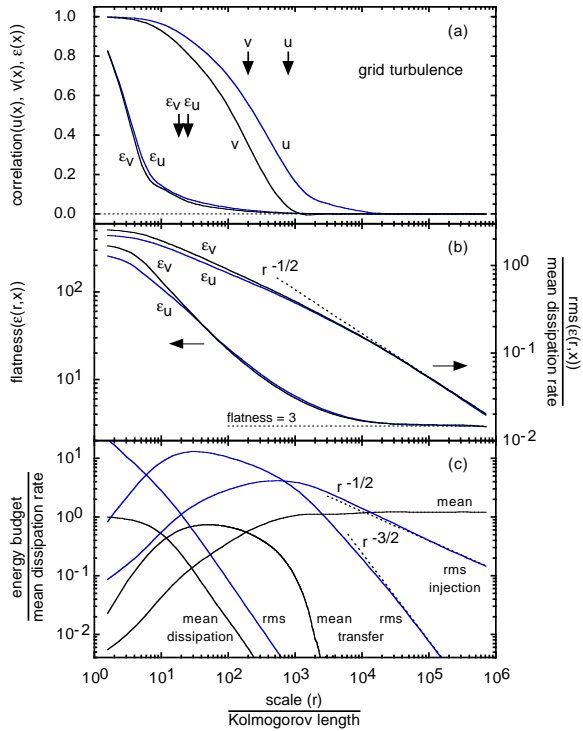}}
\caption{\label{f1} Statistics for grid turbulence as a function of scale $r$ normalized by the Kolmogorov length. (a) Correlation functions $\phi _{\varepsilon _u}(r) / \phi _{\varepsilon _u}(0)$, $\phi _{\varepsilon _v}(r) / \phi _{\varepsilon _v}(0)$, $\phi _u(r) / \phi _u(0)$, and $\phi _v(r) / \phi _v(0)$. The arrows indicate the correlation lengths $L _{\varepsilon _u}$, $L _{\varepsilon _v}$, $L_u$, and $L_v$. (b) Root-mean-square fluctuations $\sigma _{\varepsilon _u}(r)$ and $\sigma _{\varepsilon _v}(r)$. They are respectively normalized by $\langle \varepsilon _u(x) \rangle$ and $\langle \varepsilon _v(x) \rangle$. The dotted line indicates the $r^{-1/2}$ scaling. We also show flatness factors for $\varepsilon _u(r,x)$ and $\varepsilon _v(r,x)$. The dotted line indicates the Gaussian value of 3. (c) Averages and root-mean-square fluctuations of energy injection, transfer, and dissipation. They are respectively normalized by $\langle \varepsilon _v(x) \rangle$, $\langle \varepsilon _v(x) \rangle$, and $\langle \varepsilon _u(x) \rangle$. The dotted lines indicate the $r^{-1/2}$ and $r^{-3/2}$ scalings.}
\end{figure}

\subsection{Results and discussion}
\label{s3b}

The statistics of energy dissipation and relevant quantities are studied over a wide range of scale $r$. Throughout the $r$ range, statistical significance is satisfactory because our data are long.

Figure \ref{f1}(a) shows the correlation functions for the local rates of energy dissipation $\varepsilon _u(x)$ and $\varepsilon _v(x)$ and for the velocities $u(x)$ and $v(x)$. The correlation lengths are also shown. For the velocities, we defined as
\begin{subequations}
\begin{eqnarray}
\phi _u(r) = \langle u(x+r)u(x) \rangle 
\
\mbox{and}
\
L_u = \frac{ \int ^{\infty}_0 \phi _u(r) dr}{\phi _u(0)},& & \\
\phi _v(r) = \langle v(x+r)v(x) \rangle
\
\mbox{and}
\
L_v = \frac{ \int ^{\infty}_0 \phi _v(r) dr}{\phi _v(0)} .& &
\end{eqnarray}
\end{subequations}
These correlation functions and correlation lengths offer information about the scale structure of turbulence. The velocity correlations $\phi _u(r)$ and $\phi _v(r)$ exist up to the scale of largest energy-containing eddies. The correlation length $L_u$ corresponds to the mean scale of energy-containing eddies. Since the local dissipation rates belong to small scales, their correlation functions $\phi _{\varepsilon _u}(r)$ and $\phi _{\varepsilon _v}(r)$ decay fast (see Ref. \onlinecite{c03} for another interpretation). They nevertheless exist up to about the correlation length $L_u$. \cite{po97,c03}

Figure \ref{f1}(b) shows the root-mean-square fluctuations of energy dissipation $\sigma _{\varepsilon _u}(r)$ and $\sigma _{\varepsilon _v}(r)$. They are enhanced at small scales owing to intermittency. They are still comparable to the mean energy dissipation $\langle \varepsilon (x) \rangle$ at about the correlation length $L_u$. Thus, we confirm Landau's remark \cite{ll59} that the energy dissipation should significantly fluctuate over scales of energy-containing eddies. Consistent results are seen, albeit not explicitly stated, in past experimental works. \cite{po97,c03} The statistical scaling $\sigma _{\varepsilon} (r) \propto r^{-1/2}$ starts at about the scale where the velocity correlations $\phi _u(r)$ and $\phi _v(r)$ vanish, i.e., the scale of largest energy-containing eddies.

Figure \ref{f1}(b) also shows the flatness factor for the fluctuation of the energy dissipation $\varepsilon (r,x)$ around its average $\langle \varepsilon (r,x) \rangle$:
\begin{equation}
\mbox{flatness factor} 
=
\frac{\langle  [\varepsilon (r,x) - \langle \varepsilon (r,x) \rangle ]^4 \rangle }
     {\langle  [\varepsilon (r,x) - \langle \varepsilon (r,x) \rangle ]^2 \rangle ^2 } .                           
\end{equation}
The flatness factor is enhanced at small scales owing to intermittency. With an increase of scale above the correlation length $L_u$, the flatness factor approaches the Gaussian value of 3. The exact value is not achieved even at largest scales, \cite{kg02} although the discrepancy is too small to be discernible in our diagram. This is because the central limit theorem does not necessarily lead to an exactly Gaussian distribution. The discrepancy from the Gaussian value would be larger in the higher-order statistics, the reliable computation of which requires the longer data.

Except at smallest scales, the statistics for the surrogate dissipation rates $\varepsilon _u(x)$ and $\varepsilon _v(x)$ are almost the same. \cite{c03} It is thereby expected that the true dissipation rate would yield almost the same statistics.

Figure \ref{f1}(c) shows quantities in the scale-by-scale energy budget equation for decaying homogeneous isotropic turbulence: \cite{daza99,fongy00}
\begin{eqnarray}
\label{eq12}
& &
-\frac{15}{4r^5} 
 \int^{r}_0 \frac{\partial \langle \delta u(r',x)^2 \rangle}{\partial t}{r'}^4 dr'
\\ \nonumber & &
-\frac{5 \langle \delta u(r,x) ^3 \rangle}{4r}
+\frac{15 \nu}{2 r} \frac{\partial \langle \delta u(r,x) ^2 \rangle}{\partial r}
=\langle \varepsilon (x) \rangle .
\end{eqnarray}
Here $\delta u(r,x)$ is the velocity increment $u(x+r)-u(x)$. We coarse-grained the local energy injection $-15 \int^{r}_0 \partial_t ( \delta u^2 ) {r'}^4 dr' / 4r^5$, the transfer $-5 \delta u ^3/4r$, and the dissipation $15 \nu \partial _r (\delta u ^2)/2r$ over the scale $r$ and computed their averages and root-mean-square fluctuations around the averages (for the energy injection, see also Appendix). The mean energy transfer is close to the mean energy dissipation $\langle \varepsilon (x) \rangle$ in the inertial range, which extends up to the mean scale of energy-containing eddies. There the fluctuations of energy injection and transfer are comparable to each other and are several times greater than the mean energy dissipation. In particular, the fluctuation of energy injection is maximal. These fluctuations become to have the statistical scalings $r^{-1/2}$ and $r^{-3/2}$, respectively, \cite{note1} at the scale where the fluctuation of energy dissipation $\sigma _{\varepsilon}(r)$ becomes to have the statistical scaling $r^{-1/2}$.

The energy dissipation occurs at the end of the energy cascade and hence depends on the scale-by-scale energy transfer. \cite{k74} Although the mean energy transfer is downward, the local energy transfer is as often upward as downward at each scale because its fluctuation is strong [Fig. \ref{f1}(c)]. This fluctuation of energy transfer causes the large-scale fluctuation of energy dissipation. Since the former is stronger than the latter  [Figs. \ref{f1}(c) and \ref{f1}(b)], spatial mixing is at work, \cite{k74} albeit not complete. \cite{ms91} The fluctuation of energy transfer is associated with the fluctuation of energy injection. Since the local energy injection is as often negative as positive [Fig. \ref{f1}(c)], the individual energy-containing eddies as often gain as lose their energies. There is accordingly the significant fluctuation of energy dissipation over scales of energy-containing eddies [Fig. \ref{f1}(b)]. If the scale $r$ exceeds the scale of largest energy-containing eddies, the fluctuations of energy injection, transfer, and hence dissipation reflect the random and independent distribution of energy-containing eddies. They have the statistical scalings $r^{-1/2}$, $r^{-3/2}$, and $r^{-1/2}$ [Figs. \ref{f1}(c) and \ref{f1}(b)]. The fluctuation of energy dissipation is close to Gaussian [Fig. \ref{f1}(b)].

\begin{figure}[!]
\resizebox{8cm}{!}{\includegraphics*[4.5cm,7.5cm][19.5cm,26cm]{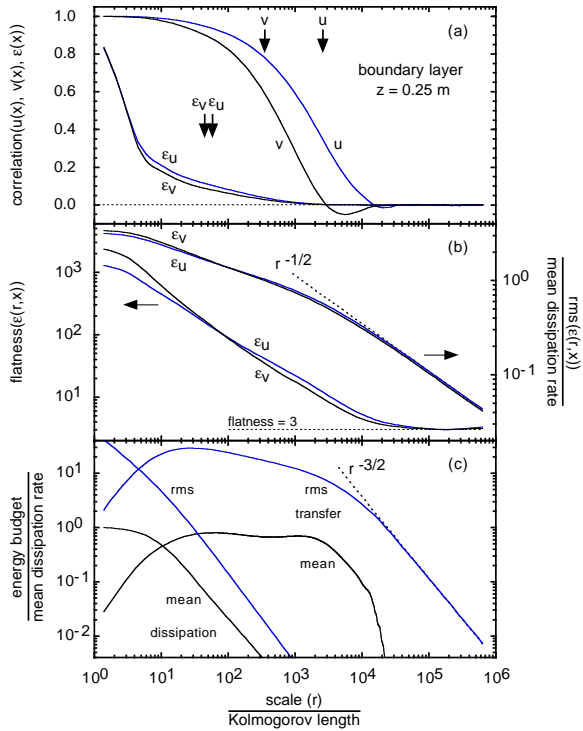}}
\caption{\label{f2} Same as in Fig. \ref{f1} but for boundary-layer turbulence at $z = 0.25\,$m.}
\end{figure}

\begin{figure}[!]
\resizebox{8cm}{!}{\includegraphics*[4.5cm,7.5cm][19.5cm,26cm]{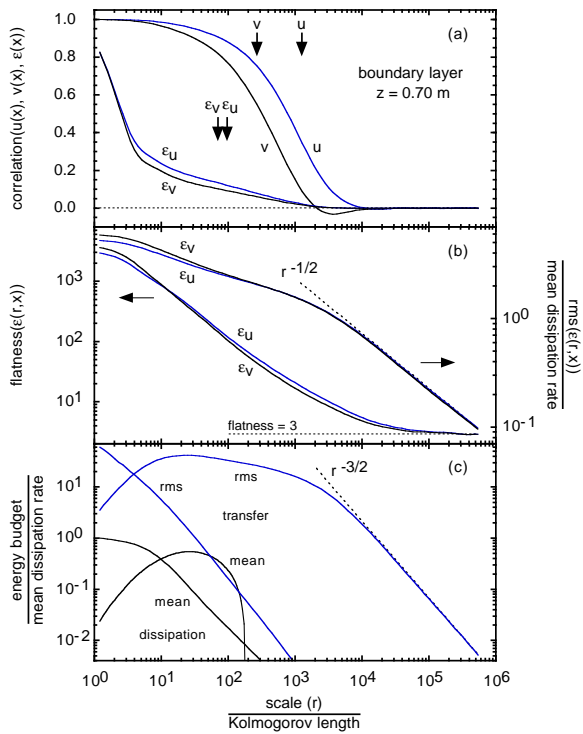}}
\caption{\label{f3} Same as in Fig. \ref{f1} but for boundary-layer turbulence at $z = 0.70\,$m.}
\end{figure}

\section{BOUNDARY-LAYER TURBULENCE}
\label{s4}

\subsection{Experiment}
\label{s4a}

The experiment was done in the same wind tunnel with the same instruments as for the grid turbulence. We describe specific features. The flow parameters are summarized in Table \ref{t1}.

Over the entire floor of the test section of the wind tunnel, we placed blocks as roughness for the boundary layer. The block size was $\delta x = 0.06$\,m, $\delta y = 0.21$\,m, and $\delta z = 0.11$\,m. The spacing of adjacent blocks was $\delta x = \delta y = 0.5$\,m. We set the incoming wind to be 10\,m\,s$^{-1}$.

The measurement positions were at $x = 12.5$ m, where the boundary layer was well developed. The 99\% thickness, i.e., the height at which the mean streamwise velocity $U$ is 99\% of its maximum value $\hat{U}$, was 0.77 m. The displacement thickness $\textstyle \int_0^{\hat{z}} (1-U/\hat{U}) dz$ was 0.20 m. Here $\hat{z}$ is the height for the velocity $\hat{U}$. \cite{note2} The log-law sublayer was at $z \simeq 0.15$--0.35\,m.

The measurement was done at $z = 0.25$ and 0.70\,m. The height $z = 0.25$\,m was in the log-law sublayer, where various eddies filled the space randomly and independently. \cite{m03a} The flatness factor $\langle v(x)^4 \rangle / \langle v(x)^2 \rangle ^2$ was close to the Gaussian value of 3. The height $z = 0.70$\,m was in the wake sublayer, where eddies did not fill the space. They intermittently passed the probe.

Since turbulence was not weak compared with the mean streamwise velocity, the $u$ component at smallest scales measured by the crossed-wire probe was contaminated with the $w$ component that is perpendicular to the two wires of the probe. The $v$ component was free from such contamination. \cite{note3} Hence we prefer the rate $\varepsilon _v(x)$ to the rate $\varepsilon _u(x)$.

Turbulence was not isotropic at large scales, but it should have been isotropic at small scales. This is because, with a decrease of scale $r$, the observed ratio between $2r \langle \delta v(r,x)^2 \rangle$ and $\partial _r (r^2 \langle \delta u(r,x)^2 \rangle )$ once becomes  the isotropic value of unity at about the correlation length $L_u$.

\subsection{Results and discussion}

Figures \ref{f2} and \ref{f3} show the same statistics as Fig. \ref{f1}, except for the energy injection that is due to large-scale shear of the boundary layer. There is no known formula to compute the local injection of energy budget in such a shear flow. \cite{daza01}  The energy transfer and dissipation are still of physical significance and their averages satisfy Eq. (\ref{eq12}) at small scales where the energy injection is negligible and the turbulence is isotropic (see Sec. \ref{s4a}).

The mean energy transfer is different. At $z = 0.25$\,m [Fig. \ref{f2}(c)], the inertial range is wide. At $z = 0.70$\,m [Fig. \ref{f3}(c)], the inertial range is narrow. There were only few eddies. Their main bodies were at $z \ll 0.70$\,m. Their small streamwise sizes at $z = 0.70$\,m correspond to the upper edge of the inertial range. The correlation length $L_u$ is relatively large because it reflects the spatial distribution of those eddies.

Despite the difference of the mean energy transfer in Figs. \ref{f2} and \ref{f3}, we find common features, albeit not exactly universal. They were also found in Fig. \ref{f1} for grid turbulence. The local rate of energy dissipation has a correlation up to about the correlation length $L_u$. There the fluctuation of energy dissipation $\sigma _{\varepsilon}(r)$ is comparable to the mean energy dissipation $\langle \varepsilon (x) \rangle$. The fluctuation of energy transfer is stronger. These fluctuations become to have the statistical scalings $r^{-1/2}$ and $r^{-3/2}$ at about the scale where the velocity correlations vanish. The fluctuation of energy dissipation becomes close to Gaussian.

Frisch \cite{f91,f95} considered that the large-scale fluctuation of energy dissipation is significant if flow parameters change in space over a scale larger than the mean scale of energy-containing eddies. An example is that turbulence does not fill the space. \cite{f95} This is the case at $z = 0.70$\,m. However, not only at $z = 0.70$\,m [Fig. \ref{f3}(b)] but also at $z = 0.25$\,m [Fig. \ref{f2}(b)], the large-scale fluctuation of energy dissipation is significant.

\begin{figure}[!]
\resizebox{8cm}{!}{\includegraphics*[4cm,6.5cm][19cm,26cm]{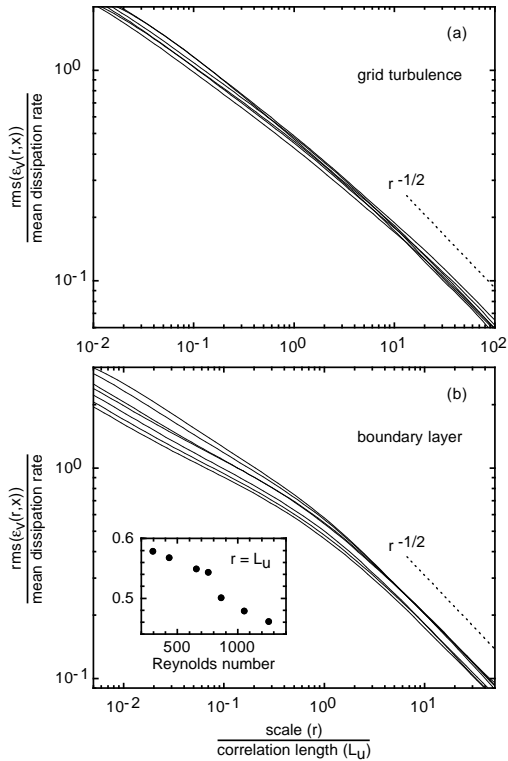}}
\caption{\label{f4} Root-mean-square fluctuation $\sigma _{\varepsilon _v}(r)$ as a function of scale $r$. The ordinate is normalized by $\langle \varepsilon _v(x) \rangle$. The abscissa is normalized by the correlation length $L_u$. The dotted line indicates the $r^{-1/2}$ scaling. (a) Grid turbulence. The data for Re$_{\lambda} = 249$ are from our present experiment. The data for Re$_{\lambda} = 105$, 165, 225, 292, and 329 are from our past experiment. \cite{m03b} (b) Boundary layer. The data for Re$_{\lambda} = 756$ are from our present experiment. The data for Re$_{\lambda} = 295$, 430, 655, 861, 1054, and 1258 are from our past experiment. \cite{m04} The inset shows the dependence on the microscale Reynolds number Re$_{\lambda}$ at $r = L_u$. }
\end{figure}

\section{DEPENDENCE ON THE REYNOLDS NUMBER}
\label{s5}

The dependence of the fluctuation of energy dissipation $\sigma _{\varepsilon _v}(r)$ on the microscale Reynolds number Re$_{\lambda}$ is studied in Fig. \ref{f4}. We use the data from our present experiments and also those from our past experiments. \cite{m03b,m04}

Figure \ref{f4}(a) shows the fluctuation $\sigma _{\varepsilon _v}(r)$ for grid turbulence at Re$_{\lambda} = 105$--329. The grid or the mean wind velocity was not the same, but the flatness factors $\langle u(x)^4 \rangle / \langle u(x)^2 \rangle ^2$ and $\langle v(x)^4 \rangle / \langle v(x)^2 \rangle ^2$ were close to the Gaussian value of 3. The ratio $\langle u^2 \rangle ^{1/2} / \langle v^2 \rangle ^{1/2}$ was close to unity. \cite{m03b}

Figure \ref{f4}(b) shows the fluctuation $\sigma _{\varepsilon _v}(r)$ for boundary-layer turbulence at Re$_{\lambda} = 295$--1258. The data were obtained in log-law sublayers at the same streamwise position $x$ over the same roughness for different incoming-wind velocities. The flatness factor $\langle v(x)^4 \rangle / \langle v(x)^2 \rangle ^2$ was close to the Gaussian value of 3. \cite{m04}

Regardless of the Reynolds number, the fluctuation $\sigma _{\varepsilon _v}(r)$ at the correlation length $L_u$ is comparable to the mean dissipation rate $\langle \varepsilon _v(x) \rangle$. The boundary-layer turbulence has a trend that the fluctuation $\sigma _{\varepsilon _v}(r)$ is smaller at a higher Reynolds number. However, at Re$_{\lambda} = 9000$ in an atmospheric boundary layer, \cite{c03} the fluctuation $\sigma _{\varepsilon _u}(r)$ at the correlation length $L_u$ is still comparable to the mean dissipation rate $\langle \varepsilon _u(x) \rangle$. The trend at Re$_{\lambda} \gg 1000$ is insignificant or absent. We could attribute the trend to spatial mixing during the energy cascade. \cite{k74,note4} The cascade depth and duration are significantly greater at a higher Reynolds number so far as it is not very high. For the grid turbulence, the trend is not discernible because the Reynolds number does not span a wide range.

\section{CONCLUDING DISCUSSION}
\label{s6}

Using grid and boundary-layer turbulence, for the first time, we experimentally confirm Landau's remark \cite{ll59} that the energy dissipation should significantly fluctuate in space over large scales. This large-scale fluctuation is caused by the large-scale fluctuation of scale-by-scale energy transfer. Although the large-scale fluctuation of energy dissipation is smoothed by spatial mixing, \cite{k74} the smoothing is not complete. \cite{ms91}

The large-scale fluctuation of energy dissipation is not exactly universal as remarked by Landau. \cite{ll59} Nevertheless, there are features that are independent of the Reynolds number and the configuration for turbulence production. The fluctuation $\sigma _{\varepsilon}(r)$ is comparable to the mean energy dissipation $\langle \varepsilon (x) \rangle$ at the correlation length of streamwise velocity $L_u$. If the scale $r$ exceeds the scale where the velocity correlations vanish, the fluctuation has the statistical scaling $\sigma _{\varepsilon}(r) \propto r^{-1/2}$ and is close to Gaussian.

The correlation length $L_u$ corresponds to the mean scale of energy-containing eddies. The velocity correlations exist up to the scale of largest energy-containing eddies. Hence, although the large-scale fluctuation of energy dissipation is caused by that of energy transfer, they are ultimately caused by individual energy-containing eddies. Exceptionally, when turbulence does not fill the space, the large-scale fluctuations are determined by the distribution of energy-containing eddies.

There are global quantities that exhibit significant temporal fluctuations. An example is the total energy injection required to sustain the K\'arm\'an flow, i.e., turbulence between two counter-rotating disks.\cite{lpf96,php03,tc05} The fluctuation is considered to be associated with the fluctuation of energy transfer. When the system size, e.g., the diameter of the two disks relative to the separation between them, is much greater than the correlation length, the fluctuation is Gaussian. The existence of such fluctuations is consistent with our results for large-scale spatial fluctuations.

\begin{figure}[!]
\resizebox{7cm}{!}{\includegraphics*[2cm,9cm][19cm,19.5cm]{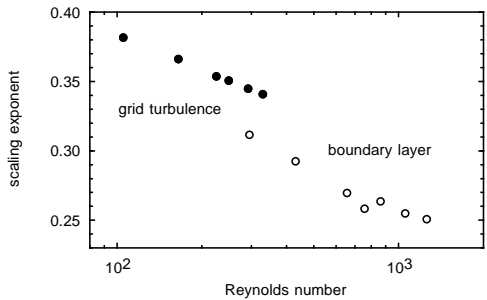}}
\caption{\label{f5} Exponent $\gamma$ for the inertial-range scaling $\sigma _{\varepsilon _v}(r) \propto r ^{-\gamma}$ as a function of the microscale Reynolds number Re$_{\lambda}$. The filled circles denote grid turbulence. The open circles denote boundary-layer turbulence. The data are the same as in Fig. \ref{f4}. }
\end{figure}

Landau \cite{ll59} also remarked that the large-scale fluctuation of energy dissipation should affect small-scale statistics in the inertial and dissipation ranges. We finally discuss this remark.

The large-scale fluctuation of energy dissipation and that of energy transfer do not affect the small-scale statistics considered by Kolmogorov, \cite{k41a} i.e., the mean energies $\langle \delta u(r,x)^2 \rangle$ and $\langle \delta v(r,x)^2 \rangle$. They are determined by the averages of energy transfer and dissipation, which satisfy the energy budget equation (\ref{eq12}) in spite of their strong fluctuations [Fig. \ref{f1}(c)--\ref{f3}(c)]. The Kolmogorov constant $\langle \delta u(r,x)^2 \rangle / (r \langle \varepsilon (x) \rangle )^{2/3}$ for the inertial range is actually universal among various flows. \cite{s95} When turbulence does not fill the space, the Kolmogorov constant is different, but the universal value is obtained if we consider the turbulence regions alone. \cite{k92} In addition, the energy spectrum for small scales is universal if it is normalized with the kinematic viscosity $\nu$ and the mean energy dissipation $\langle \varepsilon (x) \rangle$. \cite{sv94}

The large-scale fluctuation of energy dissipation and that of energy transfer affect high-order small-scale statistics, e.g., $\langle \delta u(r,x)^n \rangle$ for $n \ge 4$ and $\langle \varepsilon (r,x) ^n \rangle$ for $n \ge 2$. They represent the fluctuations of energy, energy transfer, and energy dissipation, which suffer from the large-scale fluctuations. For example, $\langle \delta u(r,x)^6 \rangle$ corresponds to the mean-square fluctuation of the energy transfer $-5 \delta u ^3/4r$. The effect of the large-scale fluctuations is not always significant. We are still able to demonstrate its existence using an inertial-range scaling in Figs. \ref{f1}--\ref{f4}:
\begin{equation}
\sigma _{\varepsilon _v}(r) \propto r ^{-\gamma},
\end{equation}
which is more significant than the familiar scalings \cite{k62,f91,f95,ms91,ss95,sa97,po97,kg02,c03,fongy00} $\langle \delta u(r,x)^n \rangle \propto r^{\zeta _n}$, $\langle \varepsilon (r,x) ^n \rangle \propto r^{\tau _n}$, and $\phi _{\varepsilon}(r) \propto r^{- \mu}$ at least in our data (see also Ref. \onlinecite{po97}). The data for Fig. \ref{f4} are used to compute the exponent $\gamma$ over the $r$ range from $30\eta$ to $0.3L_u$. Figure \ref{f5} shows the results as a function of the microscale Reynolds number Re$_{\lambda}$. The grid turbulence and the boundary-layer turbulence make up different sequences. \cite{note5} Although the flatness factor $\langle v(x)^4 \rangle / \langle v(x)^2 \rangle ^2$ was commonly close to the Gaussian value of 3, the configuration for turbulence production and hence the large-scale fluctuations are different. The exponent sequence is accordingly different. Consistent results exist for the scaling $\phi _{\varepsilon _u}(r) \propto r^{- \mu}$ and the dependence of $\delta u(r,x)$ and $\varepsilon (r,x)$ on a large-scale quantity, $u(x)$. \cite{k92,p93,ss96,sd98} Thus, since the large-scale fluctuations are not exactly  universal, the high-order small-scale statistics are not universal, although some conditional statistics might be universal. \cite{o62,ss96,sd98}

Having confirmed Landau's remark, we underline that turbulence is much more fluctuating over large scales than it is usually assumed to be. The individual energy-containing eddies as often gain as lose their energies. The local energy transfer is as often upward as downward. These significant fluctuations lead to the large-scale fluctuation of energy dissipation and also affect small-scale statistics.

\begin{acknowledgments}

U. Frisch gave us historical comments on Landau's remark. T. Gotoh pointed out that $\langle \delta u(r,x)^6 \rangle$ corresponds to the mean-square fluctuation of energy transfer. K. R. Sreenivasan pointed out that the large-scale fluctuation of energy dissipation on some level has been experimentally known but it has not been discussed with any highlighting. Our study was supported in part by the Research Promotion Fund of Doshisha University.

\end{acknowledgments}

\appendix*
\section{Taylor's hypothesis}
\label{app}

Taylor's frozen-eddy hypothesis leads to the conversion of the streamwise position $x_L^{}$ and time $t_L^{}$ in the laboratory frame into the streamwise position $x_T^{}$ and time $t_T^{}$ in a virtual reference frame (hereafter, the Taylor frame):
\begin{equation}
\label{a1}
x_T^{} = -Ut_L^{} \quad \mbox{and} \quad
t_T^{} = \frac{x_L^{}}{U}.
\end{equation}
Here $U$ is the mean streamwise velocity. The second conversion is not familiar but is necessary because eddies are not exactly frozen (see below). Then a stationary signal is converted into a streamwisely homogeneous signal. A streamwisely homogeneous signal is converted into a stationary signal.

Taylor's hypothesis is valid provided that the turbulence strength $\langle u^2 \rangle ^{1/2}/U$ is small enough. This is the case in our experiment (Table \ref{t1}). For turbulence strengths close to ours, the validity of Taylor's hypothesis was confirmed using data obtained simultaneously with two probes separated by streamwise distances. \cite{sd98}

The Taylor frame is locally identical to a reference frame moving with the mean stream. With an increase of scale, Taylor's hypothesis becomes a mere convention to describe measured temporal variations in terms of spatial variations. The signal in the Taylor frame is still expected to be consistent with some realistic turbulence. We thereby applied Taylor's hypothesis to all the scales.

Since our laboratory flows were stationary, the signals in the Taylor frame represent flows that are homogeneous in the streamwise direction. We actually obtained the scaling $\sigma _{\varepsilon}(r_T^{}) \propto r_T^{-1/2}$ and the tendency for Gaussianity expected for largest scales in homogeneous turbulence.

Since our laboratory flows were inhomogeneous in the streamwise direction, the signals in the Taylor frame represent nonstationary flows. We demonstrate this fact for the grid turbulence. If turbulence is isotropic, homogeneous, and decaying, the energy budget equation is \cite{daza99,fongy00}
\begin{equation}
\label{a2}
-\frac{15}{4r_T^5} \int^{r_T^{}}_0 \frac{\partial \langle \delta u ^2 \rangle}
                                        {\partial t_T^{}}
                                        {r_T'}^4 dr_T'
-\frac{5 \langle \delta u ^3 \rangle}{4r_T^{}}
+\frac{15 \nu}{2 r_T^{}} \frac{\partial \langle \delta u ^2 \rangle}{\partial r_T^{}}
=\langle \varepsilon \rangle .
\end{equation}
This is identical to Eq. (\ref{eq12}). The first term in the left-hand side is from the nonstationarity and represents the injection of energy budget. We estimated it from the data at $x_L^{} = 3.75$ and 4.25\,m. The other terms were estimated from the data at $x_L^{} = 4$\,m. They are compared in Fig. \ref{f1}(c). The energy-budget equation (\ref{a2}) is satisfied throughout the scales.

The above estimation appears to imply that the energy budget is supplied from the inhomogeneity of turbulence along the streamwise direction in the laboratory frame. However, once we have used Taylor's hypothesis, it is important to distinguish the laboratory frame from the Taylor frame. Grid turbulence in the Taylor frame is homogeneous in the streamwise direction and nonstationary. This nonstationarity supplies the energy budget.


\begin{thebibliography}{999}
\bibitem{note1} Since the energy transfer $-5 \delta u ^3/4r$ has the factor $r^{-1}$, its statistical scaling is $r^{-3/2}$ instead of $r^{-1/2}$.
\bibitem{note2} These thicknesses are usually defined using the incoming-wind velocity. However, since our wind tunnel was incapable of adjusting the mean streamwise pressure gradient to be zero, the mean streamwise velocity at large heights exceeded the incoming-wind velocity. We thereby used the maximum mean streamwise velocity $\hat{U} = 10.48$\,m\,s$^{-1}$ at the height $\hat{z} = 0.90$\,m.
\bibitem{note3} The two wires individually respond to all the $u$, $v$, and $w$ components. Since the measured $u$ component corresponds to the sum of the responses of the two wires, it is contaminated with the $w$ component. Since the measured $v$ component corresponds to the difference of the responses, it is free from the $w$ component.
\bibitem{note4} There remains a possibility that the trend is spurious at least partially. At a higher Reynolds number, the Kolmogorov length was smaller below the probe size. \cite{m04}
\bibitem{note5} The observed difference in the exponent sequence does not spuriously originate in large-scale anisotropy of the boundary-layer turbulence. Owing to the effect of the surface roughness, the ratio $\langle u^2 \rangle ^{1/2} / \langle v^2 \rangle ^{1/2}$ was not far from unity (1.2, Table \ref{t1} and Ref. \onlinecite{m04}).
\end{thebibliography}
\end{document}